\documentclass[conference]{IEEEtran}
\IEEEoverridecommandlockouts
\usepackage{cite}
\usepackage{amsmath,amssymb,amsfonts}
\usepackage{algorithmic}
\usepackage{url}
\usepackage{hyperref}
\usepackage[ruled,linesnumbered]{algorithm2e}
\usepackage{graphicx}
\usepackage{textcomp}
\usepackage{xcolor}
\def\BibTeX{{\rm B\kern-.05em{\sc i\kern-.025em b}\kern-.08em
    T\kern-.1667em\lower.7ex\hbox{E}\kern-.125emX}}
\begin{document}

\title{Using Adamic-Adar Index Algorithm to Predict Volunteer Collaboration: Less is More\\
}

\author{\IEEEauthorblockN{Chao WU$^1$}
\and
\IEEEauthorblockN{Peng CHEN$^2$}
\and
\IEEEauthorblockN{Baiqiao YIN$^3$}
\and
\IEEEauthorblockN{Zijuan LIN$^4$}
\and
\IEEEauthorblockN{Chen JIANG$^5$}
\and
\IEEEauthorblockN{Di YU$^6$}
\and
\IEEEauthorblockN{Changhong ZOU$^7$}
\and
\IEEEauthorblockN{Chunwang LUI$^8$}

\thanks{$^1$School of Life Sciences, The Chinese University of Hong Kong, HKSAR, \href{mailto://wuchao@link.cuhk.edu.hk}{wuchao@link.cuhk.edu.hk}}
\thanks{$^2$School of Science, University of Hong Kong, HKSAR, \href{mailto://pchen38@outlook.com}{pchen38@outlook.com}}
\thanks{$^3$School of intelligent system engineering, Sun Yat-sen University, China, \href{mailto://FS13942366949@163.com}{FS13942366949@163.com}}
\thanks{$^4$School of Science and Engineering, Chinese University of Hong Kong(Shenzhen), China, \href{mailto://zijuanlin@link.cuhk.edu.cn}{zijuanlin@link.cuhk.edu.cn}}
\thanks{$^5$School of Business, Hong Kong University of Science and Technology, HKSAR, \href{mailto://cjiangak@connect.ust.hk}{cjiangak@connect.ust.hk}}
\thanks{$^6$Institute for Ocean Engineering, Tsinghua Shenzhen International Gradual School, China, \href{mailto://yud22@mails.tsinghua.edu.cn}{yud22@mails.tsinghua.edu.cn}}
\thanks{$^7$Institute of biopharmaceutical and health engineering, Tsinghua Shenzhen International Graduate School, China, \href{mailto://1518577034@qq.com}{1518577034@qq.com}}
\thanks{$^8$Division of Humanities and Social Sciences, Tsinghua Shenzhen International Graduate School, China, \href{mailto://lvjh23@mails.tsinghua.edu.cn}{lvjh23@mails.tsinghua.edu.cn}}

}

\maketitle

\begin{abstract}
Social networks exhibit a complex graph-like structure due to the uncertainty surrounding potential collaborations among participants. Machine learning algorithms possess generic outstanding performance in multiple real-world prediction tasks. However, whether machine learning algorithms outperform specific algorithms designed for graph link prediction remains unknown to us. To address this issue, the Adamic-Adar Index (AAI), Jaccard Coefficient(JC) and common neighbour centrality(CNC) as representatives of graph-specific algorithms were applied to predict potential collaborations, utilizing data from volunteer activities during the Covid-19 pandemic in Shenzhen city, along with the classical machine learning algorithms such as random forest, support vector machine, and gradient boosting as single predictors and components of ensemble learning. This paper introduces that the AAI algorithm outperformed the traditional JC and CNC, and other machine learning algorithms in analyzing graph node attributes for this task. 
\end{abstract}


\section{Introduction}
Social networks are complicated systems and this characteristic brings challenges for machine learning scientists to predict collaboration activities in real-world scenarios, such as high dimensionality and data sparsity issues\cite{perozzi2014deepwalk}. Due to its complexity, link prediction requires us to predict the possibility of two nodes in a graph generated by social network feature extraction to form links. Not like other prediction tasks which can be sorted as Euclidean data-oriented tasks, graphs have their own structural plasticity, which also enhances the difficulties in mimicking their nature\cite{bronstein2017geometric}.

Generally, graphs are fundamental structures involving paired objects and they are related to some extent. This kind of relationship is called ‘edge’ and can be illustrated as binary format(directed or undirected) in a graph\cite{li2019graph}. Link prediction is a fundamental question in network analysis and it requires us to predict the score between two nodes’ relationships in certain networks. To understand the dynamics of a social network, link prediction data can facilitate this kind of requirement and is widely applicable\cite{zhang2018link}. To be more specific, similarity of common neighbors-based algorithms are well-known approaches to tackle link prediction tasks.

Recently, many attempts have been explored to solve the link prediction problem, especially for real-world tasks. For example, Mohan et al\cite{mohan2017scalable} proposed an advanced community detection system that utilized link prediction algorithms and Bulk Synchronous Parallel programming model. They meant to cater a single system environment and increase the scalability of link prediction algorithms in community detection. Local indices are one of the most straightforward approaches, considering the number of neighbor nodes and the degree of neighbor to calculate the similarity score in link prediction. For instance, local similarity indices get a greater prediction accuracy with Area Under Curve(AUC) scores up to 0.99, by combining topological structure with the computed similarity score\cite{xu2017similarity} or with additional community information\cite{sun2017improved}. However, since local indices only consider neighbor nodes, some interesting and potential links may be missed. It can be further improved by developing a parallel algorithm for the similarity calculation task. In addition, Muniz et al\cite{muniz2018combining} proposed a combination of global similarity indices and content-based measures in their work to improve the performance of link prediction. As compared to local index approaches, global indices identify all direct and indirect paths that are interesting to be included in the similarity score. But in the context of link prediction in large networks, global similarity indices are time-consuming and computationally complex due to high dimensionality of networks. In the future, there remains a need for more scalable global approach to handle link prediction in a distributed environment. Nevertheless, Quasi-local approaches use additional topological information as global indices do. They compute the score based on nodes with a path distance of no more than two, which is similar to local index approaches and achieve higher prediction accuracy than local approaches as they consider additional topological information while acquiring lower computation complexity\cite{ozcan2016temporal}. At the same time, neural network approaches also play pivotal roles in link prediction tasks. For example, Kagan et al\cite{kagan2018generic} present a novel unsupervised two-layered meta-classifier that can detect irregular vertices in complex networks solely by utilizing topology-based features, which is based on the reasoning that a vertex with many improbable links has a higher likelihood of being anomalous. The authors use the Random Forest algorithm to construct the link prediction algorithm for their training sets . They chose the Random Forest algorithm because previous link prediction studies demonstrated that, in most cases, it performs better than other classification algorithms at predicting links. The proposed anomaly detection algorithm demonstrates very low false positive rates, on average 0.006, in all of the tested scenarios. In addition, the algorithm generated especially good results, with an average AUC of 0.99 and a false positive rate(FPR) of 0.021 in semi-simulated graphs, i.e., real-world networks with injected anomalous vertices. 

Among them, LINE (Large-scale Information Network Embedding) model is the most related example for collaboration prediction in social networks\cite{tang2015line}. Specifically, the authors utilized random forest algorithm to solve this issue. The robustness of random forest is well known in modern biological and bioinformatic studies due to its excellent fitness with laboratory data. However, whether it performs better than graph-specific algorithms such as JC and AAI\cite{liben2003link} on link prediction tasks is still unknown yet important.

This paper introduces the comparisons between machine learning and graph-specific algorithms on link prediction tasks. First, we compared the performance of JC, AAI, and CNC \cite{ahmad2020missing} on volunteer collaboration dataset. And we evaluated the performance of several machine learning algorithms as single predictors, components of ensemble learning and Graph Convolutional Networks (GCN) \cite{kipf2016semi}. We found that AAI outperformed other two graph-specific algorithms and complex machine learning algorithms lead to overfitting on this task.

\section{Preliminaries}
\subsection{Jaccard coefficient}

Suppose we have a social network $G = \langle V, E\rangle$ in which each edge in this network $e = \langle u,v \rangle$ belongs to $E$, which represents that nodes $u$ and $v$ have interaction. Jaccard coefficient (JC) is a similarity measurement between two sets of nodes in each network. It is defined as the ratio of the intersection size to the union size of the two node sets. Where $(u,v)$ is a pair of nodes, $\Gamma(u)$ denotes the number of neighbors of node $u$, and $c$ denotes the coefficient. The formula of JC is defined as \eqref{jc}.
\begin{equation}
    \label{jc}
    c=\frac{\left|\Gamma\left(u\right)\cap\Gamma\left(v\right)\right|}{\left|\Gamma\left(u\right)\cup\Gamma\left(v\right)\right|}
\end{equation}

\subsection{Adamic-Adar Index}
Adamic-Adar Index (AAI) \cite{liben2003link} is a local similarity measurement and gives a higher weight to shared neighbors that are less common in certain networks and a lower weight to more commonly shared neighbors respectively. AAI of node $u$ and $v$ is defined as \eqref{aai}. Here $\Gamma$ follows the same definition of \eqref{jc}, $w$ is the common neighbor of $u$ and $v$, the resulting index is $I$.
\begin{equation}
    \label{aai}
    I=\sum_{w\in\Gamma\left(u\right)\cap\Gamma\left(v\right)}\frac{1}{\log{\left|\Gamma\left(w\right)\right|}}
\end{equation}

\subsection{Common Neighbor Centrality}
Common neighbor centrality(CNC) is designed to measure the importance of a node in certain network with respect to its common neighbors shared with other nodes. The formula of CNC is defined as \eqref{cnc}. Here the resulting score is $s$, $\alpha \in [0,1]$ is a hyper-parameter, $N$ denotes the total number of nodes in the graph and $d_{uv}$  denotes the shortest distance between $u$ and $v$.
\begin{equation}
    \label{cnc}
    s=\alpha\cdot\left(\left|\Gamma\left(u\right)\cap\Gamma\left(v\right)\right|\right)+\left(1-\alpha\right)\cdot\frac{N}{d_{uv}}
\end{equation}

\subsection{Random Forest}
Random forest (RF) is a classical machine learning algorithm that combines multiple tree predictors and significantly improves the predictive performance on both classification and regression tasks. Random forest is an ensemble method and obtains well-known robustness on real-world tasks. For each tree in the forest algorithm, it performs bootstrap sampling and randomly select features to build decision trees to make a prediction for input vector. Specifically, input vector will be passed through each decision tree and the predictions will be aggregated using a majority vote for classification tasks. A formal definition of RF is as \cite{breiman2001random}.

\subsection{Support Vector Machine }
Support vector machine (SVM) is designed as a powerful and robust tool to find a hyperplane as a boundary to maximize the margin and separate given data into different classes. SVM algorithm converts data to feature vectors and select hyperplane along with the optimization of margin, and finally project data points onto the hyperplane to decide which side they belong to. The pesudocode of SVM is shown as Algorithm.~\ref{svm}.

\begin{algorithm}[htbp]
\label{svm}
\SetAlgoLined
\KwData{Training dataset $D$ with $N$ samples and $M$ features}
\KwResult{Optimal hyperplane parameters $\mathbf{w}$ and $b$}
\SetKwInOut{Parameter}{Parameter}
\Parameter{Regularization parameter $C$, Learning rate $\eta$}
\BlankLine
Initialize $\mathbf{w}$ and $b$ to zeros\;
\While{Not converged}{
    \For{each sample $(\mathbf{x}_i, y_i)$ in $D$}{
        \If{$y_i(\mathbf{w} \cdot \mathbf{x}_i + b) < 1$}{
            Update $\mathbf{w} \leftarrow \mathbf{w} + \eta (y_i \mathbf{x}_i - 2C \mathbf{w})$\;
            Update $b \leftarrow b + \eta y_i$\;
        }
        \Else{
            Update $\mathbf{w} \leftarrow \mathbf{w} - \eta (2C \mathbf{w})$\;
        }
    }
}
\caption{SVM Algorithm}
\end{algorithm}

\subsection{Gradient boosting}
Gradient boosting (GB) is another tree-based strong learner for classification tasks. Begin with an initial model, GB algorithm calculates its errors and fits a new decision tree to the errors, and learns the missed patterns due to the initial model. After several times of repeats, GB algorithm adds new trees to the model until the errors are minimized. The pesudocode of GB is shown as Algorithm.~\ref{gb}.

\begin{algorithm}[htbp]
\label{gb}
\SetAlgoLined
\KwData{Training dataset $D$ with $N$ samples and $M$ features}
\KwResult{Ensemble of weak learners $F(x)$}
\SetKwInOut{Parameter}{Parameter}
\Parameter{Number of weak learners $T$, Learning rate $\eta$}
\BlankLine
Initialize $F_0(x)$ to a constant value, e.g., the mean of the target variable\;
\For{$t \leftarrow 1$ \KwTo $T$}{
    Calculate the negative gradient $r_t = - \frac{\partial}{\partial F(x)} L(y, F(x))$\;
    Fit a weak learner $h_t(x)$ to the negative gradient $r_t$ with respect to the training data $D$\;
    Update the ensemble $F_t(x) = F_{t-1}(x) + \eta h_t(x)$\;
}
\caption{GB Algorithm}
\end{algorithm}

\subsection{Stacking Ensemble Learning for Link Prediction}
Ensemble learning is a machine learning strategy that combines multiple models to improve prediction accuracy and robustness over a single model. It can be applied to both supervised and unsupervised learning scenarios, such as classification, regression, clustering, or anomaly detection. It includes methods such as Bagging, used in Random Forests, \cite{breiman2001random}, Boosting, as in AdaBoost, \cite{freund1997decision}, Gradient Boosting, \cite{friedman2001greedy}, XGBoost, \cite{chen2016xgboost}, LightGBM, \cite{ke2017lightgbm}, CatBoost, \cite{prokhorenkova2018catboost}, and Stacking or Stacked Generalization, \cite{wolpert1992stacked}. It also includes methods such as Voting, Averaging, or Blending that combine the predictions of different models using simple rules or weights. These techniques leverage the strengths of multiple models to achieve better performance.

In our research, we drew inspiration from the work of Wang, Jiao, and Wang \cite{wang2022link}, who proposed a novel stacking ensemble framework for link prediction in complex networks. Their method successfully integrated global, local, and quasi-local topological information. They employed Recursive Feature Elimination based on Random Forest (RF-RFE) to select network-related structural features and constructed a two-level stacking ensemble model involving various machine learning methods for link prediction. In their model, the base layer consisted of three base classifiers, namely, Logistic Regression (LR), Gradient Boosting Decision Tree (GBDT), and XGBoost, and then integrated the outputs of these models with the XGBoost model at the upper layer. Their research conducted extensive experiments on six networks, and the results showed that their method performed excellently in prediction results and applicability robustness compared with existing methods.

In our study, we simplified our problem to a binary classification task, where '0' indicates no collaboration and '1' indicates collaboration between volunteers. We selected Support Vector Machines (SVM), Gradient Boosting Decision Trees (GBDT), and Random Forest (RF) as our base classifiers due to their proven performance in classification tasks. These base classifiers were then integrated using Random Forest as the meta-classifier in our stacking ensemble. The performance of this ensemble model will be discussed later in this paper.

\subsection{GCNGAN}
Graph Convolutional Networks (GCN) and Generative Adversarial Networks (GAN) are two powerful machine learning frameworks that have been successfully applied in various domains. GCN, introduced by Kipf and Welling \cite{kipf2017semisupervised}, is an efficient variant of convolutional neural networks that operates directly on graphs, making it particularly effective on graph-structured data. On the other hand, GAN is known for its ability to generate new data instances that resemble the training data. A novel method that combines these two frameworks is GCNGAN (Graph Convolutional Networks Generative Adversarial Networks) \cite{zhang2018gaan}. In GCNGAN, GCN is used to capture the graph structure and node features, while GAN is employed to generate new graph structures that resemble the original graph. This combination has proven useful for tasks like graph generation and link prediction.

\section{Experimental Results}
\subsection{Graph-specific Algorithm Performance}
Three common parameters used in link prediction for graphs are:
\begin{itemize}
    \item Adamic Adar index: A measure of the similarity between two nodes based on the number of common neighbors they have, with more weight given to neighbors that are less connected overall.
    \item Jaccard coefficient: A measure of the similarity between two nodes based on the ratio of the number of neighbors they have in common to the total number of neighbors they have.
\item  Common neighbor coefficient: A measure of the similarity between two nodes based on the number of neighbors they have in common.
\end{itemize}

These parameters can be used to predict links between nodes in a graph based on their similarity to one another.

Three metrics were used to evaluate parameter performance in our problem: AUC, Precision at K, and Precision-Recall curve.

AUC measures classifier performance by indicating how well it distinguishes between positive and negative examples. Precision at K measures precision of the top K predicted links, while the Area Under Precision-Recall curve measures classifier performance by measuring its ability to identify positive examples.

These metrics comprehensively evaluate parameter performance in our problem. AUC and Precision at K measure classifier ability to distinguish between positive and negative examples, while the Precision-Recall curve measures its ability to identify positive examples in a dataset.

First, we calculate the CNC, JC, and AAI metrics on the training, testing, and validation sets. Then, we use AUC, P@K, and AUPR to evaluate their performance, and the result is shown as Table.~\ref{perf}.

\begin{table}[ht]
\label{perf}
\centering
\caption{Performance Evaluation of CNC, JC, and AAI Metrics on Different Datasets}
\begin{tabular}{|l|l|l|l|}
\hline
Metric & Training Set & Testing Set & Validation Set \\ 
\hline
CNC & & & \\ 
AUC-ROC & 0.6469 & 0.6473 & 0.6832 \\ 
P@K=50 & 1 & 1 & 1 \\ 
AUPR & 0.7728 & 0.7703 & 0.8167 \\ 
\hline
JC & & & \\ 
AUC-ROC & 0.7478 & 0.8349 & 0.8323 \\ 
P@K=50 & 0.8 & 0.9 & 0.8 \\ 
AUPR & 0.7858 & 0.8472 & 0.8414 \\ 
\hline
AAI & & & \\ 
AUC-ROC & 0.8571 & 0.8897 & 0.8878 \\ 
P@K=50 & 1 & 1 & 0.9064 \\ 
AUPR & 0.9099 & 0.9107 & 0.9096 \\ 
\hline
\end{tabular}
\end{table}
The three link prediction parameters evaluated in the study displayed varying levels of performance when assessed using the AUC-ROC, P@K, and AUPR metrics. The Jaccard Coefficient (JC) parameter, for instance, demonstrated relatively poorer performance when evaluated using the P@K metric, while the Common Neighbor Criterion (CNC) parameter was found to have a comparatively lower performance on the AUPR metric. On the other hand, the Adamic-Adar Index (AAI) parameter was found to have the best overall performance when all the metrics were taken into consideration. These findings suggest that the choice of the link prediction parameter could have a significant impact on the quality of the predictions. Therefore, it is essential to carefully evaluate the performance of each parameter using multiple metrics to ensure the best possible results.

\subsection{Single Machine Learning Algorithm Performance}
We examined the performance of widely used machine learning algorithms and made comparison between them. First, we examined the predictive performance of the RandomForestRegressor algorithm and got a model acc of only 47.76\%, which is much lower than we expected. We believe that this may be due to the small number of features provided by the data, randomforest cannot use information entropy for regression very clearly, and it can be improved in the future by performing more detailed feature engineering. Next, we also examined the Support Vector Machine (SVM),GradientBoosting, and Random Forest, and their performance is summarized in the following table.

\begin{table}[ht]
\centering
\caption{Performance Comparison of SVM, GradientBoosting, and Random Forest Algorithms}
\begin{tabular}{|l|l|l|l|}
\hline
Metric & SVM & GradientBoosting & Random Forest \\ 
\hline
0-precision & 0.81 & 0.81 & 0.79 \\ 
0-recall & 0.74 & 0.76 & 0.84 \\ 
0-f1\_score & 0.78 & 0.78 & 0.81 \\ 
1-precision & 0.82 & 0.83 & 0.87 \\ 
1-recall & 0.87 & 0.86 & 0.84 \\ 
1-f1\_score & 0.84 & 0.85 & 0.85 \\ 
accuracy & 0.82 & 0.82 & 0.84 \\ 
\hline
\end{tabular}
\end{table}

We find that classical individual machine learning models are more biased to predict the results as category 1, resulting in higher recall for category 1. Moreover, the tree model outperforms the others, and among the tree models, Random Forest is the best performer. But so the overall accuracy of the models are not high, so more in-depth algorithms can be explored.

\subsection{Ensemble Learning Algorithm Performance}
We applied an ensemble learning approach from and found that it has weak prediction ability in terms of the recruited dataset. The ensemble learning algorithm’s performances on test dataset and validation dataset are shown.

\begin{table}[ht]
\centering
\caption{Performance of Ensemble Learning Algorithm on Test and Validation Datasets}
\begin{tabular}{|l|l|l|}
\hline
Metric & Test Dataset & Validation Dataset \\ 
\hline
0-precision & 0.79 & 0.80 \\ 
0-recall & 0.84 & 0.84 \\ 
0-f1\_score & 0.81 & 0.82 \\ 
1-precision & 0.87 & 0.87 \\ 
1-recall & 0.83 & 0.84 \\ 
1-f1\_score & 0.85 & 0.85 \\ 
accuracy & 0.84 & 0.84 \\ 
\hline
\end{tabular}
\end{table}

\begin{figure}[ht]
\centering
\includegraphics[width=0.45\textwidth]{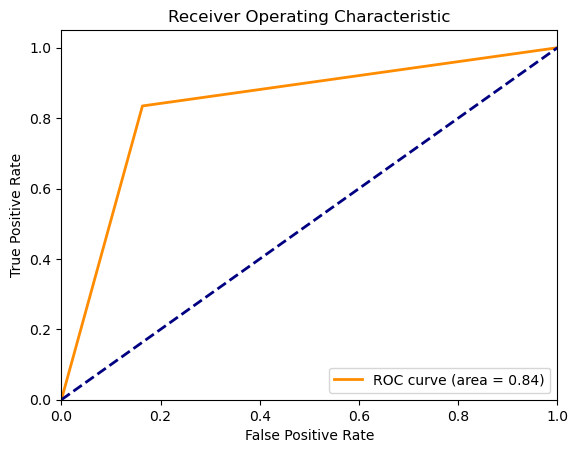} 
\caption{Stacking Ensemble Framework's ROC curve on test dataset}
\label{fig:test_dataset_ROC}
\end{figure}

\begin{figure}[ht]
\centering
\includegraphics[width=0.45\textwidth]{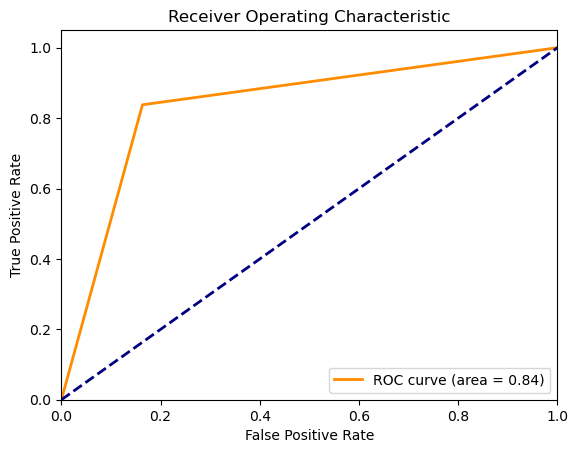} 
\caption{Stacking Ensemble Framework's ROC curve on validation dataset}
\label{fig:validation_dataset_ROC}
\end{figure}
Observing the results, it can be found that the integrated framework performs better than a single machine learning method in all aspects, and various metrics have been improved to a greater or lesser extent, indicating that the strategy we applied is effective.

\subsection{GCNGAN}
We also tried a graph neural network-based approach: using GCNGAN\cite{8737631}, a temporal link prediction model with recognizable weights, to predict links temporally. This is done by counting the number of times two people work together in the 19 months from 2020.02 to 2021.08 based on whether they appear in the same task or not in the test set of volunteers, and using the number of times in each month as the weight of the subgraph of the cooperation situation in each month, and through the weighted graphs of the 19 months, we use the GCNGAN model to make predictions in the temporal dimension for the next five months The principle of the GCNGAN model is that when learning distribution features with GCN in the spatial dimension, the generator and discriminator of GAN are utilized to fit the pattern of temporal changes in the temporal dimension. However, the accuracy of the prediction is not very high, only 59\%. We believe this is because the 19-month time span used for training is still too short to learn robust temporal features.

\section{Conclusion}
We have compared the performance of multiple algorithms on graph link prediction tasks and found that AAI algorithm outperforms others in volunteer collaboration prediction task. And AAI algorithm has preferred predictive performance on positive link generations. This insight can be utilized and expanded in similar real-world prediction tasks.
\bibliographystyle{IEEEtran}
\bibliography{ref}

\end{document}